# OpenExtract: Automated Data Extraction for Systematic Reviews in Health

Jim ACHTERBERG [a,1], Bram VAN DIJK[a], Jing MENG[b], Saif Ul ISLAM[c],
Gregory EPIPHANIOU[c], Carsten MAPLE[c], Xuefei DING[d],
Theodoros N. ARVANITIS[d], Simon BROUWER[e], Marcel HAAS[a] and
Marco SPRUIT[a,b]

[a] *Leiden University Medical Center, The Netherlands*
[b] *Leiden Institute of Advanced Computer Science, Leiden University, The Netherlands*
[c] *University of Warwick, UK*
[d] *School of Engineering, University of Birmingham, UK*
[e] *Syntho BV, The Netherlands*

ORCiD ID: Jim Achterberg https://orcid.org/0009-0000-9589-7831, Bram van Dijk https://orcid.org/0009-0002-9176-1608, Saif Ul Islam https://orcid.org/0000-0002-9546-4195, Gregory Epiphaniou https://orcid.org/0000-0003-1054-6368, Carsten Maple https://orcid.org/0000-0002-4715-212X, Xuefei Ding https://orcid.org/0009-0001-3073-9514, Theodoros N. Arvanitis https://orcid.org/0000-0001-5473-135X, Simon Brouwer https://orcid.org/0000-0002-0956-0851, Marcel Haas https://orcid.org/0000-0003-2581-8370, Marco Spruit https://orcid.org/0000-0002-9237-221X

**Abstract.** This study presents OpenExtract, an open-source pipeline for automated data extraction in large-scale systematic literature reviews. The pipeline queries large language models (LLMs) to predict data entries based on relevant sections of scientific articles. To test the efficacy of OpenExtract, we apply it to a systematic literature review in digital health and compare its outputs with those of human researchers. OpenExtract achieves precision and recall scores of > 0.8 in this task, indicating that it can be effective at extracting data automatically and efficiently. OpenExtract: https://github.com/JimAchterbergLUMC/OpenExtract.

## 1. Introduction

The number of published scientific papers is surging [1], which renders data extraction in systematic literature reviews (SLRs) a labour-intensive task. Especially when the sought-after information is specific, researchers must resort to manual inspection of numerous articles. Here, we present OpenExtract, an open-source pipeline for SLR data extraction assisted by large language models (LLMs). It supports researchers in exploring broad research areas of interest for the review without sacrificing granularity in the data extraction process. In this way, our tool helps obtain deeper insights from SLRs, even when the topic is broad and yields many papers.

[1] Corresponding Author: Jim Achterberg, j.l.achterberg@lumc.nl

LLMs handle language processing tasks that require specific types of knowledge and diverse inputs well, given the size and variability of their training data [2]. So, they may allow (partly) automating labour-intensive tasks in SLRs. As useful machine-learning tools for literature screening exist (e.g., ASReview [3]), abstracting each step to an LLM leaves researchers with too little control over the scope of the SLR and the process. Hence, we focus only on utilizing LLMs for the data extraction stage. Previous work has demonstrated some success in addressing this issue [4,5]. OpenExtract is especially powerful in its flexibility, by allowing to use virtually any text encoder and LLM through Hugging Face and OpenRouter. This can be useful for i) academic purposes, by benchmarking various LLMs, ii) financial purposes, by selecting the most cost-efficient alternative, and iii) domain adaptation, by adapting the pipeline to different domains by selecting appropriately fine-tuned text encoders and LLMs.

In the remainder of this work, we first present OpenExtract, then introduce the digital health SLR used to evaluate the pipeline, and finally present results on its efficacy compared with human data extraction.

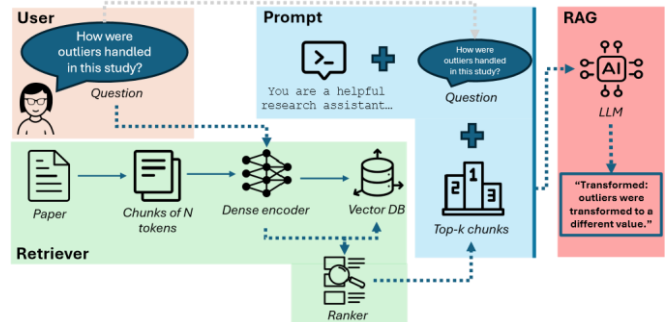

**Figure 1.** OpenExtract: a pipeline for LLM-assisted data extraction in SLRs.

## 2. Methods

### 2.1. OpenExtract

Essentially, we develop a retrieval augmented generation (RAG) pipeline, where we first identify the article chunks most relevant to a given data entry, then feed those chunks into an LLM to select the most appropriate answer(s) from a set of labels. Figure 1 provides an overview of the pipeline. RAG adds additional task-relevant texts to the context of an LLM [2]. We use it to encourage the LLM to base its answers on the actual article, rather than relying on more general knowledge from its training data.

We divide each article into 1000-token segments, with 500 overlapping tokens between segments. We use a BERT model finetuned on PubMed[2] as a domain-specific embedding model [6] to encode chunks and identify the top-3 most relevant chunks per data entry (i.e., user query) based on cosine similarity between their embeddings and those of the article chunks. The top 3 most relevant chunks provide additional context for the LLM.

---

[2] https://huggingface.co/NeuML/pubmedbert-base-embeddings

BERT models truncate inputs to a maximum of 512 tokens. To enable relatively long, continuous token streams as input to the LLM, we chunk articles into 1000-token segments, allowing 500-token overlap. This approach, together with BERT truncation, serves as a sliding-window mechanism, ensuring that each part of the article is evaluated for relevance.

The LLM receives article chunks and a data entry in its input prompt and is tasked with predicting the answer from a list of output labels. We use the following prompts, primarily aimed at encouraging the LLM to answer only from article chunks, and output answers in a structured format (JSON):

**Base:** *You are a helpful and meticulous research assistant who answers questions about study details carefully and adequately from context chunks of provided research papers.* Formatting: *Carefully read the QUESTION, ANSWERS, AND CONTEXT. If one or more options are correct, return ONLY a JSON array of the correct option IDs, e.g., ["A","C"]. Return the IDs exactly as provided in ANSWERS (case-sensitive). Do NOT include any text besides the JSON array.* **Question (example):** *"Which data type is used in this study?"* **Answers (example):** *["Tabular", "Time-series", "Images", "Text", "Video", "Audio", "Multi-modal"]* **Context:** contains the top-3 most relevant article chunks.

We use the OpenRouter interface[3] to query the LLMs, which allows us to test a variety of models through a single API. We select the following LLMs based on their open-weights nature, low costs and varying number of parameters: DeepSeek V3.1 (671B parameters) [7], and Qwen2.5 Instruct (72B and 7B parameter versions) [8].

*2.2. Systematic Literature Review in Digital Health*

As OpenExtract is meant to simplify large-scale data extraction in SLRs, we test it on a broad topic likely to yield many relevant articles, while still extracting detailed information from each article. We consider the following research question: *Which data analysis techniques are suitable for the various types of digital health applications?* We prepared 15 data entries and corresponding potential output labels to aid in answering this question. These data entries consider topics such as the type of digital health application, the number of study participants, prediction models, and evaluation metrics. A detailed overview of data entries and output labels can be found in the GitHub repository[4].

We searched Web of Science, IEEE Xplore, PubMed, and ACM Digital Library for all dates up until June 2025 for journal articles and conference proceedings using the following search query:

("data analysis" OR "machine learning" OR "artificial intelligence" OR "deep learning" OR "predictive analytics" OR "statistical analysis" OR "data mining")
AND

---

[3] https://openrouter.ai/
[4] https://github.com/JimAchterbergLUMC/OpenExtract

("digital health" OR "mHealth" OR "eHealth" OR "telemedicine")

Initial search resulted in 7323 articles. We employ ASReview to aid in title/abstract screening. ASReview employs active learning to recommend the most relevant articles for the researcher to consider next [3]. This way, whenever $k$ irrelevant articles are suggested in a row, it can be expected that other articles are also irrelevant, and the screening stage can be stopped early. As inclusion criteria, we consider only articles on digital health applications that i) employ predictive analytics and/or machine learning, and ii) are tested in a real-life medical setting. We stop screening when ASReview suggests 10 irrelevant articles in a row. This screening stage yields 249 relevant articles.

## 3. Results

The outputs of OpenExtract were compared with the data as extracted by two human researchers (the first two authors). We randomly selected 50 papers to evaluate the pipeline. In this paper, we present results from the evaluation of the first 10 papers in this random subset, corresponding to 150 data points (10 papers, 15 data entries each).

Table 1 presents the inter-rater reliability (measured by Cohen's κ) between the two human researchers and our pipeline across three LLMs. Results show that researchers tend to agree much more with each other than with LLMs. This is to be expected, as discussions regarding the interpretation of data entries could not be fully captured in the LLM prompts. Also, the LLMs tend to agree more with each other than with humans, hinting at potential bias of the LLMs.

**Table 1.** Inter-rater reliability (Cohen's $\kappa$) between researchers and language models in data extraction.

|              | Researcher 1 | Researcher 2 | DeepSeek | Qwen (72B) | Qwen (7B) |
|---|---|---|---|---|---|
| **Researcher 1** | 1.000 | 0.925 | 0.763 | 0.784 | 0.512 |
| **Researcher 2** |       | 1.000 | 0.749 | 0.751 | 0.496 |
| **DeepSeek**     |       |       | 1.000 | 0.852 | 0.606 |
| **Qwen (72B)**   |       |       |       | 1.000 | 0.618 |
| **Qwen (7B)**    |       |       |       |       | 1.000 |

To provide an absolute measure of the pipeline's effectiveness, we report precision and recall scores in Table 2. Here, we noticed that for many questions in our SLR (and other SLRs alike), there is no single ground truth in data extraction. We therefore consider a True Positive (TP) to be any answer from an LLM that was provided by one of the researchers, and a False Positive (FP) to be an answer from an LLM that was provided by *neither of the researchers*. Likewise, a False Negative (FN) is an answer provided by *both of the researchers* but missed by the LLM. With these statistics, we can compute the scores as $Precision = \frac{TP}{TP+FP}$ and $Recall = \frac{TP}{TP+FN}$.

**Table 2.** Precision and Recall of language models in data extraction.

|           | DeepSeek | Qwen (72B) | Qwen (7B) |
|---|---|---|---|
| **Precision** | 0.820 | 0.846 | 0.624 |
| **Recall**    | 0.820 | 0.813 | 0.564 |

Interestingly, Table 2 shows that DeepSeek V3.1 does not outperform the much smaller Qwen 2.5 Instruct (72B). We expect that this is because the task at hand mainly concerns in-context knowledge retrieval, and the vast amount of general knowledge embedded in large LLMs such as DeepSeek V3.1 is not necessarily beneficial here. However, as we move to much smaller models, such as Qwen 2.5 Instruct (7B), we observe that the smaller parameter size substantially harms performance.

## 4. Discussion and Conclusion

This work presents OpenExtract, an open-source RAG pipeline to perform automated data extraction in large-scale SLRs. Initial results from an SLR in digital health were promising, with precision and recall > 0.8, compared with those obtained by human researchers. Moreover, OpenExtract can be used for any structured data extraction task from text, including many use cases beyond SLRs.

Still, we also acknowledge some limitations. Although the SLR served an illustrative purpose, the search query is very broad and does not use MeSH terms. Regarding the pipeline, information embedded in figures and tables is currently missed because we only parse text. Future improvements can also consider parsing figures and tables using multimodal foundation models.

## 5. Acknowledgements

This work is co-funded by the HORIZON.2.1 - Health Programme of the European Commission, Grant Agreement number: 101095661 - Innovative applications of assessment and assurance of data and synthetic data for regulatory decision support (INSAFEDARE).

## References

[1] Park M, Leahey E, Funk RJ. Papers and patents are becoming less disruptive over time. Nature. 2023;613(7942):138-44.
[2] Yang J, Jin H, Tang R, Han X, Feng Q, Jiang H, et al. Harnessing the power of llms in practice: A survey on chatgpt and beyond. ACM Transactions on Knowledge Discovery from Data. 2024;18(6):1-32.
[3] Van De Schoot R, De Bruin J, Schram R, Zahedi P, De Boer J, Weijdema F, et al. An open source machine learning framework for efficient and transparent systematic reviews. Nature machine intelligence. 2021;3(2):125-33.
[4] Dagdelen J, Dunn A, Lee S, Walker N, Rosen AS, Ceder G, Persson KA, Jain A. Structured information extraction from scientific text with large language models. Nature Communications. 2024;15(1):1418.
[5] Khan MA, Ayub U, Naqvi SAA, Khakwani KZR, Sipra ZbR, Raina A, et al. Collaborative large language models for automated data extraction in living systematic reviews. Journal of the American Medical Informatics Association. 2025;32(4):638-47.
[6] Gu Y, Tinn R, Cheng H, Lucas M, Usuyama N, Liu X, et al. Domain-specific language model pretraining for biomedical natural language processing. ACM Transactions on Computing for Healthcare (HEALTH). 2021;3(1):1-23.
[7] DeepSeek-AI, Liu A, Feng B, Xue B, Wang B, Wu B, et al.. DeepSeek-V3 Technical Report; 2025. Available from: https://arxiv.org/abs/2412.19437.
[8] Yang A, Yang B, Zhang B, Hui B, Zheng B, Yu B, et al.. Qwen2.5 Technical Report; 2025. Available from: https://arxiv.org/abs/2412.15115.